\documentstyle[amssymb,epsfig]{mn2e}

\title[The radial mass density profile of the Galactic Halo Globular Cluster System]
{On the origin of the radial mass density profile of the Galactic halo 
Globular Cluster System}
\author[G.~Parmentier \& E.~K.~Grebel]{Genevi\`eve Parmentier 
\thanks{E-mail: gparm@ast.cam.ac.uk} \& Eva K.~Grebel \\
Astronomical Institute, University of Basel, 7 Venusstrasse, 
CH-4102 Binningen, Switzerland \\}
\begin{document}

\date{Accepted .... Received ... ; in original form ...}

\pagerange{\pageref{firstpage}--\pageref{lastpage}} \pubyear{2004}

\maketitle

\label{firstpage}
 
\begin{abstract}
We investigate what may be the origin of the presently observed  spatial 
distribution of the mass of the Galactic Old Halo globular cluster system.
We propose its radial mass density profile 
to be a relic of the distribution of the cold baryonic material
in the protoGalaxy.  Assuming that this one arises from the profile of 
the whole protoGalaxy minus the contribution of the dark matter (and a small
contribution of the hot gas by which the protoglobular clouds were bound),  
we show that the mass distributions around the Galactic centre of 
this cold gas and of the Old Halo agree satisfactorily.  In order to 
demonstrate our hypothesis even more conclusively, we simulate the 
evolution with time, up to an age of 15\,Gyr, of a putative globular 
cluster system  whose initial mass distribution in
the Galactic halo follows the profile of the cold protogalactic 
gas.  We show that beyond a galactocentric distance of order 2 to 3\,kpc, 
the initial shape of such a mass density profile is preserved 
inspite of the complete destruction of some globular clusters and the partial 
evaporation of some others.  This result is almost independent of the 
choice of the initial mass function for the globular clusters, which is 
still ill-determined.  The shape of these evolved cluster system mass 
density profiles also agree with the 
presently observed profile of the Old Halo globular cluster system,
thus strengthening our hypothesis.  Our result might suggest that the 
flattening shown by the Old Halo mass density profile at short distance 
from the Galactic centre is, at least partly, of primordial origin.

\end{abstract}

\begin{keywords}
globular clusters: general -- Galaxy: halo -- Galaxy: formation 
\end{keywords}

\section{Introduction}

Globular Clusters (GC) are thought to be the oldest bound stellar systems
in our Galaxy.  Their study provides therefore valuable information 
about the early Galactic evolution.  In this respect, a major problem 
is that we do not know whether what we presently observe is still 
representative of the initial conditions and, thus, a fossil imprint
of the formation process, or whether the initial conditions have been 
wiped out by a 15\,Gyr long evolution within the tidal fields of
the Milky Way.
Modelling the dynamical evolution of the Galactic Globular Cluster 
System (GCS) is thus of great interest as it helps us to go back
in time to the earliest stages of the cluster system and to disentangle
the formation and evolutionary fingerprints (see, e.g., 
Gnedin \& Ostriker 1997, Baumgardt 1998, Vesperini 1998, Fall \& Zhang 2001).
The GCs most vulnerable to evaporation and disruption are the low-mass 
clusters located at small galactocentric distance.  As a result,
the evolution with time of a GCS is markedly determined by the initial 
distribution of the GCs in space around the Galactic centre as well 
as by their initial mass spectrum. 

As for the presently observed spatial distribution of 
the Galactic halo GCs, it is centrally concentrated with the density
varying as $D^{-3.5}$ ($D$ is the Galactocentric distance)
over most of the halo (Zinn 1985).   In the inner 3-4\,kpc, 
the distribution flattens to something closer to an $D^{-2}$ dependence.
As a result, the overall distribution is conveniently described by a 
power-law with a core (see Section 2).  The observed central flattening 
probably arises from a combination of several 
effects: our failure to discover some GCs in the heavily absorbed
central regions of the Galaxy,
distance errors, and the real flattening of the distribution.  
It is still unclear whether such a flattening is of primordial origin
and reflects the initial spatial distribution of the system, 
or whether it
has been completely determined by evolutionary processes.  The latter
are especially effective at small galactocentric distances where the 
GC relaxation time is short, causing the disruption of some GCs and 
the partial evaporation of some others.  More generally,
as far as we are aware of, the shape of the whole spatial distribution
of the halo GCS has received no explanation.  
   
The aim of this paper is to present a possible model accounting for
the distribution around the Galactic centre of the mass content of the
halo GCS, that is, the variation with galactocentric distance of the mass
density of the halo GCS.  We will consider solely the Old Halo (OH) GCs,
excluding from our analysis the Younger Halo (YH) subsystem.   This one is
made of GCs suspected of having been accreted and is thus of limited relevance
to the earliest stages of the main body of our Galaxy (see, e.g., Zinn 1993,
van den Bergh 1993).  
A common trend of different scenarios for the formation of GCs 
is to assume that their gaseous progenitor clouds, whatever their
size ($\sim 10^6\,{\rm M}_{\odot}$, e.g., Fall \& Rees 1985; 
$\sim 10^9\,{\rm M}_{\odot}$, e.g., Harris \& Pudritz 1994), are
embedded in a hot and tenuous background at the virial temperature.
We build on this picture, splitting the protoGalaxy in a set of three
components: a gaseous medium consisting of a hot and a cold phases
(in the form of a collection of cold and dense clouds, the gaseous 
precursors of the halo GCs, pressure-bound by a hot medium), 
and a dark matter corona.
We investigate whether the mass density profile of the halo GCS may 
trace the profile of the cold phase, that is, of the cold baryonic matter
which was available to star formation some 15\,Gyr ago. \\
 
The outline of the paper is as follows.  In Section 2, we build the 
radial mass density profile of the OH and obtain
new fits of a power-law with a core, with parameter values appropriate to
this subsystem of GCs only.  We also summarize the evidence from the
literature following which mass-related quantities may be better probes 
to the initial conditions than number related quantities.  
In Section 3, we present our hypothesis regarding the shape of the initial
mass density profile of the OH GCS and we compare the OH profile 
obtained in Section 2 with our suggested model, that is, with 
the radial density profile of the cold protogalactic material.  
In Section 4, we simulate the evolution with time of a GCS whose
initial mass density profile mirrors the one of the cold protogalactic 
gas, and this for various initial GC mass spectra.  We compare the 
spatial mass distributions of these evolved systems with the presently
observed spatial mass distribution of the OH.
Finally, our conclusions are presented in Section 5. 

\section{Radial distribution of the halo GCS mass}

The observed radial distribution of the Galactic GCS (i.e., the number of GCs
per unit volume in space as a function of Galactocentric distance $D$) 
is often parametrized by a simple power law with a core:
\begin{equation}
\rho (D) = \rho _0 \left(1+\frac{D}{D_c} \right)^{-\gamma}\;.
\label{eq:pl_core}
\end{equation}
Realizing a ``chi-by-eye'' fitting of this type of curve on the observed  
spatial distribution of the GCS, Djorgovski \& Meylan (1994) found good 
matches for $\gamma \sim$ 3.5-4.0 and $D_c \sim$ 0.5-2\,kpc, the steepest 
slope being associated with the largest core.
This approach is purely empirical however and is not meant to imply any 
physical meaning of the distribution given by equation (\ref{eq:pl_core}).  

Both the number and the mass density profiles of the halo 
([Fe/H] $<$ --0.8) GCS can be described by equation (\ref{eq:pl_core}),
as both profiles are indistinguishable in shape. 
However, as argued by McLaughlin (1999), the spatial distribution of the mass
is likely to be a better estimate of the initial conditions than
its number counterpart.  Dynamical evolution targetting mostly 
low-mass clusters and these ones accounting for a limited fraction
of the GCS mass (see below and Section \ref{sec:evol_rho_D}), 
the decrease with time of the total GCS mass is much
slower than the decrease of the total number of GCs.  

This interesting property comes from how the shape of the initial 
mass spectrum of the halo GCs may have looked like.
In our Galaxy, the luminosity function of the halo GCs 
\footnote{In what follows, we adopt the nomenclature of 
McLaughlin \& Pudritz (1996).  We call mass/luminosity {\sl spectrum} 
the number of objects per {\sl linear} luminosity/mass interval, 
${\rm d}N/{\rm d}m$ or ${\rm d}N/{\rm d}L$,
while we refer to the mass/luminosity {\sl function} to describe
the number of objects per {\sl logarithmic} luminosity/mass interval, 
${\rm d}N/{\rm d~log}~m$ or ${\rm d}N/{\rm d~log}~L$. }
(the number of GCs per unit absolute magnitude, which is proportional to 
the number of objects per logarithmic mass interval)
is bell-shaped and usually fitted with a gaussian.
However, the underlying mass spectrum (i.e., the number of objects per 
linear mass interval) is well fitted by a two-index power-law, with exponents 
$\sim -2$ and  $\sim -0.2$ above and below $\sim 1.5 \times 
10^5$ M$_{\odot}$, respectively (McLaughlin 1994).   
The peak of the gaussian magnitude function in fact coincides with the
cluster mass at which the slope of the mass spectrum changes.
The slope of the high mass regime 
is reminiscent of what is observed in interacting and merging galaxies 
(see, e.g., Whitmore \& Schweizer 1995, Whitmore et al.~2002) where
systems of young GCs show well defined power-law with slopes ranging between
--1.8 and --2
for their luminosity spectrum (but see the discussion in Section 
\ref{sec:evol_rho_D}).  This thus suggests that the initial mass spectrum
of the halo GCS may have been itself a single-power law.
Numerous studies of GCS dynamical evolution modelling have shown that a 
Hubble time long evolution turns such an initial spectrum into the 
presently observed one (e.g., Baumgardt 1998, Vesperini 1998, 
Fall \& Zhang 2001).  In fact, low-mass clusters being the most 
vulnerable to evaporation and disruption, the GC mass spectrum gets 
severly depleted below a turnover of $\sim 1.5 \times 10^5$ M$_{\odot}$, 
leading to a much shallower mass spectrum (i.e., slope $\simeq-0.2$) 
in the low-mass regime.  Pal~5 constitutes a striking example of a low-mass
GC currently dissolved by the Galactic tidal fields 
(Odenkirchen et al., 2001; Dehnen et al., 2004).

\begin{table*}
\begin{center}
\caption[]{OH sample}
\label{tab:OH}
\begin{tabular}{ l l l l l l l l l l } \hline 
NGC288  & NGC5286 & NGC5986 & NGC6218 & NGC6293 & HP1     & NGC6522 & NGC6638 & NGC6752 \\ 
Pal2    & NGC5466 & NGC6093 & NGC6235 & NGC6341 & NGC6362 & NGC6535 & NGC6652 & NGC6779 \\ 
NGC1904 & NGC5634 & NGC6121 & NGC6254 & NGC6325 & NGC6402 & NGC6540 & NGC6656 & NGC6809 \\ 
NGC2298 & NGC5694 & NGC6101 & Pal15   & NGC6333 & NGC6401 & NGC6544 & NGC6681 & NGC6864 \\ 
NGC4372 & NGC5824 & NGC6144 & NGC6266 & NGC6355 & NGC6397 & NGC6541 & NGC6712 & NGC7078 \\ 
NGC4833 & NGC5897 & NGC6139 & NGC6273 & IC1257  & NGC6426 & NGC6558 & NGC6717 & NGC7089 \\
NGC5024 & NGC5904 & NGC6171 & NGC6284 & NGC6366 & NGC6453 & NGC6569 & NGC6723 & NGC7099 \\ 
NGC5053 & NGC5946 & NGC6205 & NGC6287 & Ter4    & NGC6517 & NGC6626 & NGC6749 & NGC7492 \\ \hline
\end{tabular}
\end{center}
\end{table*}   

Assuming that most of the GCs more massive than the turnover 
are spared by the dynamical evolution, McLaughlin (1999) compares 
the GCS initial and final mass spectra (i.e., the single power-law with 
the two-index power-law) and derives useful formulae 
to estimate the fraction of surviving clusters, both in term of mass and 
numbers (his equations 4-7).  His results show
that mass-related quantities are reasonably preserved by a Hubble time 
long evolution, even though low-mass GCs are disrupted in large numbers.  
Considering the specific case of the Milky Way (power-law slopes of --0.2 
and of around --1.8 to --2.0 for the low and high-mass regimes respectively, 
a turn-off mass 
of $1.5 \times 10^5 {\rm M}_{\odot}$, and lower and upper mass limits  
of $10^4 {\rm M}_{\odot}$ and $10^6 {\rm M}_{\odot}$, respectively),
the application of his formulae shows that the initial mass of the 
GCS is decreased by 40 per cent, while the fraction
of surviving clusters is 16 per cent.  In case of a lower limit for the 
cluster initial mass range, say $10^3 {\rm M}_{\odot}$,
the contrast between the decreases in mass and number is even
more striking, i.e., the survivors still represent 44 per cent of the initial
total mass but 2 per cent of the initial number only.  McLaughlin (1999) 's 
formulae thus illustrate that, compared to number related quantities,
mass related quantities are less markedly affected by a 
15\,Gyr long evolution.   As a result, the radial mass density profile 
of the halo GCS may be considered as a reasonably reliable estimator of 
their initial distribution around the Galactic centre.
We will make this point more quantitative in Section 
\ref{sec:evol_rho_D} and show that such an hypothesis is indeed robust. \\

In this paper, we are interested in understanding the origin of
the initial spatial distribution of the halo GCS mass within the Milky Way,  
which we approximate by the presently observed mass density profile
at this stage of the discussion.  We do not consider the more
metal-rich, presumably second generation, bulge/disc GCs 
([Fe/H] $\geq -0.8$).  Also, the halo ([Fe/H] $< -0.8$) subsystem
itself could be divided   
into two groups, traditionally referred to as the OH and 
the YH (see, e.g., Zinn 1993, Van den Bergh 1993, 
Mackey \& Gilmore 2004).  Evidence supporting the 
existence of such two distinct halo subsystems have been accumulating 
over the past years.  OH and YH GCs show differences in 
horizontal branch morphology, age, kinematics, spatial distribution 
(see Parmentier et al.~2000, their Section 2, for a review), as well as 
differences in the distribution of their core radius (Mackey \& 
Gilmore 2004).  The properties of the 
OH group are consistent with the majority of its members having been 
formed "in situ" during the large-scale collapse of the protogalactic cloud,
as envisioned by Eggen, Lynden-Bell \& Sandage (1962). On the other hand, 
the YH GCs are not native to the Galaxy, probably having been formed in 
external dwarf galaxies and afterwards accreted into the outer halo 
while their host galaxies were being swallowed by the Milky Way,
as suggested by Searle \& Zinn (1978). The current accretion 
of the Sagittarius dwarf galaxy and of its small GCS is the smoking 
gun of this process.  As we are interested in the mass density profile 
of the GCS which formed within the original potential well of the 
Galaxy, we restrict our attention to the OH GCs.
This OH/YH division has already been proven most fruitful as the OH GCS
shows a metallicity gradient and obeys a mass-metallicity relation, 
features predicted by simple self-enrichment models (Parmentier 
et al.~2000 and Parmentier \& Gilmore 2001, respectively), while
the whole halo GCS (OH+YH) does not. 

Lists of OH and YH GCs are provided in Lee et al.~(1994) and 
Da Costa \& Armandroff (1995).  With respect to these, we have
made two slight changes however.  In our present OH sample, 
we ignore NGC~2419. Although coeval with the inner halo (Harris et
al., 1997; Salaris \& weiss 2002), this GC is located at a  
galactocentric distance of order 90\,kpc and is thus unlikely to belong
to the main body of the Galaxy.  Moreover, van den Bergh \& Mackey (2004)
show that NGC~2419 and $\omega$ Cen on the one hand, and the other halo GCs
on the other hand, are at different locii in a half-light radius vs 
absolute visual magnitude diagram.  They thus suggest that, 
as $\omega$ Cen (which we also exclude from our sample), NGC~2419
might be the tidally stripped core of a former dwarf spheroidal
galaxy.  An efficient tidal stripping would however require NGC~2419 
to cross the inner Galactic regions.  Unfortunately, its orbit 
is still ill-determined.  Also, unlike $\omega$ Cen, there is
no evidence for a metallicity spread among the cluster giants.  If
NGC~2419 is actually the remnant of a former dwarf galaxy, the parent
galaxy might, like the Ursa Minor dwarf galaxy (van den Bergh 2000), have
produced a single generation of metal-poor stars.  Bearing these
caveats in mind, we thus note that the main peculiarity of NGC~2419
with respect to the bulk of the OH is its large galactocentric distance. 
Neglecting NGC~2419, the OH is thoroughly contained within 
D$\lesssim$40\,kpc.  Additionally, we have moved the cluster NGC~6864 
from the YH group to the OH group.
According to the former (1999) edition of the GC McMaster Catalog
(Harris 1996), the metallicity and the horizontal branch ratio (HBR) 
of NGC~6864 are --1.32 and --0.42, respectively.  The updated (2003) 
values being [Fe/H]=--1.16 and HBR=--0.07, the
location of this cluster in the [Fe/H] vs HBR diagram shows that
it is more likely a member of the OH group rather than a YH GC.
Finally, about 25 halo GCs of the Harris Catalogue still miss HB index 
measurements.  Using recently published colour-magnitude diagrams, 
(e.g., Piotto et al., 2002), Mackey \& Gilmore (2004) have sorted
most of these yet undefined GCs.  All the GCs to which Mackey \& Gilmore
(2004) have assigned an OH membership have been added to our OH
sample.  This one is presented in Table \ref{tab:OH}.    \\

Before proceeding further, we derive estimates for the parameters 
$\gamma$ and $D_c$ appropriate for the sole OH, as previous fits refer
to either the whole Galactic GCS or the whole halo system, thus ignoring
their heterogeneity.  We fit
\begin{equation}
Log_{10} ~\rho (D) = Log_{10} ~\rho _0 - \gamma ~Log_{10}  \left(1+\frac{D}{D_c} \right)
\label{eq:log_pl_core}
\end{equation}
to the observed OH mass density distribution through a 
Levenberg-Marquardt algorithm (Press et al.~1992).
Our source for the galactocentric distances $D$ and the absolute 
visual magnitudes M$_v$ is the McMaster database compiled and maintained 
by Harris (1996, updated February 2003).
Cluster absolute visual magnitudes have been turned into luminous
mass estimates by assuming a constant mass-to-light ratio m/L$_v$=2.35.  
This value corresponds to the average of the mass-to-light ratios of the
halo GCs for which Pryor \& Meylan (1993) derived dynamical mass estimates. 
The mean m/L$_v$ is almost independent of the halo group considered, 
that is, OH or OH+YH. 

The OH mass profile is derived by binning the data with two different 
bin sizes: $\Delta {\rm log} D = 0.1~ {\rm and}~ 0.2$ ($D$ is in kpc), 
corresponding to 16 and 8 points, respectively.
As for the size of the error bars, a Poissonian error on the number of GCs 
in each bin is combined with a fixed error on the mass-to-light ratio.
In fact, not all GCs show the same mass-to-light ratios, the standard
deviation in the Pryor \& Meylan (1993) compilation being of order 
$\sigma _{{\rm Log}(M/L_v)} = 0.17$.  \\  

\begin{figure}
\begin{center}
\epsfig{figure=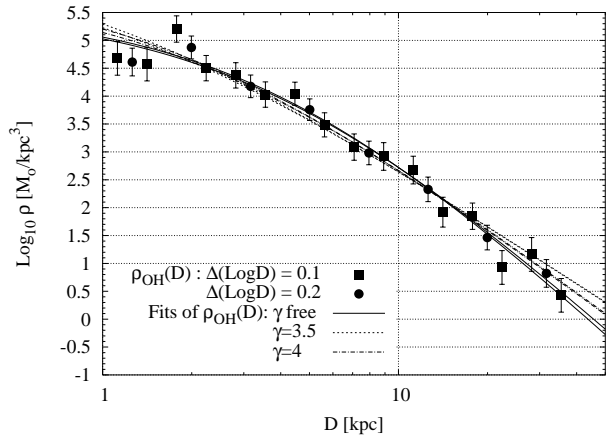, width=\linewidth}
\caption{Fits of a power-law with a core on the observed mass 
density profile of the OH subsystem.  The different curves 
correspond to the values of the parameters as given in Table
\ref{tab:fit_pl_core}} 
\label{fig:rho_OH_fits}
\end{center}
\end{figure}

Results of the fitting procedure are presented in Table \ref{tab:fit_pl_core}
and superposed to the observed distribution in Fig.~\ref{fig:rho_OH_fits}.
For each fit, we also give the $\chi ^2$ and the incomplete gamma 
function $Q(\nu /2, \chi ^2/2)$ ($\nu$ is the number of degrees of 
freedom) which provides a quantitative measure for the goodness-of-fit 
of the model \footnote{We remind the reader that a $Q$ value of 0.1 or larger
indicates a satisfactory agreement between the model and the data} 
(Press\ et al.~1992). 
Keeping all three parameters free, we obtain a slope much steeper 
($\gamma \simeq -5$) than suggested by previous works 
($\gamma \simeq -3.5 {\rm ~~to} -4.0$; Zinn 1985, Djorgovski \& Meylan 1994).  

\begin{table}
\begin{center}
\caption[]{Results of fitting a power-law with a core on the observed mass 
density profile of the OH subsystem.}
\label{tab:fit_pl_core}
\begin{tabular}{ c c c c r c} \hline 
$\Delta {\rm Log} D$ & $Log_{10}(\rho _0)$ & $D_c$  & $-\gamma$ & $\chi ^2$~ &  Q \\ \hline
0.1   & 5.6 $\pm$ 0.3&  3.7 $\pm$ 2.0 & --5.1 $\pm$ 0.9 & 11.7 &0.55 \\ 
0.2   & 5.6 $\pm$ 0.4&  3.5 $\pm$ 2.4 & --4.9 $\pm$ 1.1 &  3.8 &0.58 \\ \hline 
      &              &               &$\gamma$ imposed &     &      \\ \hline  
0.1   & 6.4 $\pm$ 0.4&  0.9 $\pm$ 0.3 & --3.5          & 17.7 & 0.22 \\ 
0.2   & 6.3 $\pm$ 0.5&  1.0 $\pm$ 0.4 & --3.5          &  7.0 & 0.32 \\ \hline
0.1   & 6.0 $\pm$ 0.3&  1.7 $\pm$ 0.4 & --4.0          & 13.9 & 0.46 \\ 
0.2   & 5.9 $\pm$ 0.3&  1.9 $\pm$ 0.5 & --4.0          &  4.8 & 0.57 \\ \hline 
\end{tabular}
\end{center}
\end{table}   

Our slope is coupled with a rather large core 
($D_c \simeq 3.6$\,kpc), however.  In fact, functions like equation
(1) or (2) show a core-slope degeneracy in the sense that two distinct 
fits can provide a satisfactory agreement with the data provided that 
a steeper slope is associated with a larger core.  This is also reflected by
the large error bars for $D_c$ and $\gamma$.
Therefore, when describing a density profile through equation 
(\ref{eq:pl_core}) or (\ref{eq:log_pl_core}), 
it is important to quote the slope as well as the core length. 
We have thus performed additional fits in which the value of the exponent 
is set to values previously quoted in the literature, i.e., --3.5 and --4. 
Such shallower slopes again provide satisfactory agreement with the data, 
as indicated in Table \ref{tab:fit_pl_core}.  With respect to
Djorgovski \& Meylan (1994), who fitted the whole Galactic GCS 
(i.e., disc plus halo), core sizes are not markedly different.

\section{Origin of the OH GCS mass density profile}
As already mentioned, the description of the OH GCS by equation 
(\ref{eq:pl_core}) or (\ref{eq:log_pl_core}) is purely empirical and 
does not imply any physical meaning.  As far as we are aware of, 
no explanation has been put forward to explain the shape of the 
radial mass density profile $\rho _{OH}(D)$.  In this paper,
we propose that $\rho _{OH}(D)$ is tracing the mass distribution 
of the gas which, some 15\,Gyr ago, was available to the process 
of star formation.  \\

In the dark matter potential well, the protogalactic gas probably 
settles into a two-phase medium consisting of (1) a hot and tenuous
gas at the virial temperature, $ T_{hot} = 1.7 \times 10^6$\,K and (2) 
a collection of much colder self-gravitating clouds, 
pressure-bound by the hot medium in which they are embedded.
These cold and dense clouds are often considered as the formation sites 
of the halo stars and GCs.  In fact, this description of the protoGalaxy 
is encountered in widely different pictures for halo GC formation.
For instance, in the frame of the Galaxy formation model relying on a 
monolithic collapse, Fall \& Rees (1985) suggested that the development 
of such a  two-phase medium is promoted by a thermal instability.  
On the other hand, adopting the hierarchical picture of Galaxy formation
as the framework of their model, Harris \& Pudritz (1994) proposed that 
the formation of GCs took place in the densest parts of large protogalactic 
fragments, of mass $10^8 {\rm -} 10^9 {\rm M}_{\odot}$ (i.e., their 
Super Giant Molecular Clouds, SGMC), these SGMCs being,
as in the Fall \& Rees (1985) model, embedded in a hot background 
at the virial temperature.  \\

\begin{figure}
\begin{center}
\epsfig{figure=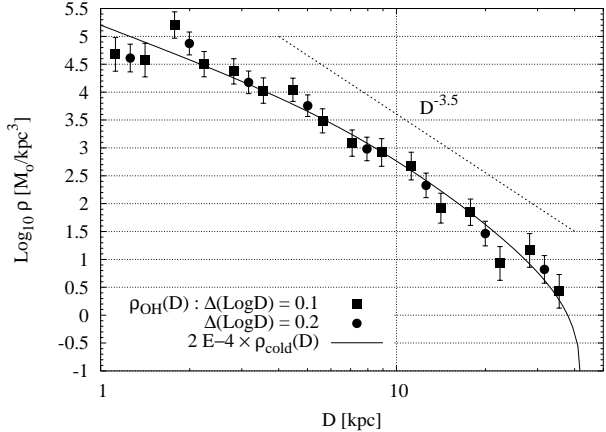, width=\linewidth}
\caption{Fit of the model $\rho _{cold}(D)$ (equation 8) to the 
presently observed mass density profile of the OH subsystem.  
The best (in the $\chi  ^2$ sense) fit is obtained when a downwards 
scaling of $2 \times 10^{-4}$ is applied to $\rho _{cold}(D)$}  
\label{fig:rho_OH_mod}
\end{center}
\end{figure}

We thus assume that the protogalaxy includes three components: a dark 
matter corona and a set of star forming clouds embedded in a hot and diffuse 
background, and we write:
\begin{equation}
\rho _{Gal}(D)=\rho _{cold}(D) + \rho _{hot}(D) + \rho _{DM}(D)\;.
\label{eq:rho_tot}
\end{equation}
In this equation, $\rho _{Gal}$, $\rho _{cold}$, $\rho _{hot}$ 
and $\rho _{DM}$ are the mass densities of the protoGalaxy, the cold 
phase, the hot phase and the dark matter, respectively.

The density profile $\rho _{Gal}$ of the protoGalaxy is conveniently 
described as a singular isothermal sphere:
\begin{equation}
\rho _{Gal} = \frac{V_c^2}{4 \pi G} \, \frac{1}{D^2}\,,
\label{eq:rho_Gal}
\end{equation}
where $V_c$ is the circular velocity of the gas in the 
dark matter potential well of the Galaxy and G is the gravitational 
constant.  Since we are mainly interested in the proto-Milky Way, 
we adopt $V_c$=220\,km.s$^{-1}$.

Although physically unmotivated, we adopt the usual description of the
spatial distribution of the dark matter mass in the Galaxy, that is:
\begin{equation}
\rho _{DM}(D)=\frac{\rho _{0DM}}{1+\left( \frac{D}{D_{DM}}\right) ^2}\;.
\label{eq:rho_DM}
\end{equation} 
As for the central density and the softening length, we adopt the values 
of the Caldwell \& Ostriker (1981) model: 
$\rho _{0DM}=13.72 \times 10^{6}\,{\rm M}_{\odot}.{\rm kpc}^{-3}$ and 
$D_{DM}=7.8151$\,kpc.

Regarding the hot gas confining the protoglobular clouds, its pressure 
profile is expected to scale as $D^{-2}$ (e.g., Murray \& Lin 1992,
Harris \& Pudritz 1994):

\begin{equation}
P _{hot}(D)=\frac{1.25 \times 10^{-9}}{D_{kpc}^2} \, {\rm dyne.cm^{-2}}\,, 
\label{eq:press_hot}
\end{equation} 
where the coefficient comes from Murray \& Lin (1992). 

Combining equation (\ref{eq:press_hot}) with the virial temperature $T_{hot}$,
we obtain the mass density profile of the hot tenuous gas:

\begin{equation}
\rho _{hot}(D)=\frac{\rho_{0hot}}{D^2} 
              = \frac{78.38}{D_{kpc}^2} \times 10^6 {\rm M_{\odot}.kpc^{-3}}\;. 
\label{eq:rho_hot}
\end{equation} 

The combination of equations (\ref{eq:rho_tot}), (\ref{eq:rho_Gal}), 
(\ref{eq:rho_DM}) and (\ref{eq:rho_hot}) provides us with an estimate 
of the radial mass density profile of the cold gas, that is, 
the gas out of which halo GCs presumably formed:
\begin{equation}
\rho _{cold}(D)= \left( \frac{890.7}{D^2} - \frac{78.4}{D^2} 
- \frac{13.7}{1+\left(\frac{D}{7.8}\right)^2}  \right)
\times 10^6 {\rm M_{\odot}.kpc^{-3}}\;. 
\label{eq:rho_cold}
\end{equation} 
   
\begin{figure}
\begin{center}
\epsfig{figure=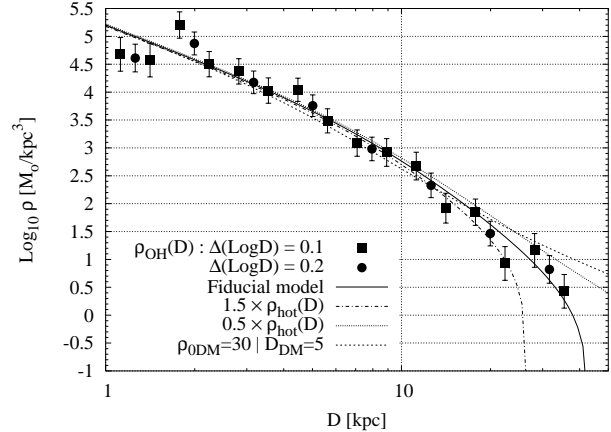, width=\linewidth}
\caption{Comparison between the radial mass density profile of the OH 
subsystem and the model mass density $\rho _{cold}(D)$,
considering different values for the coefficients in equation (8)}   
\label{fig:rho_OH_modbis}
\end{center}
\end{figure}

Figure \ref{fig:rho_OH_mod} displays equation (\ref{eq:rho_cold}), 
as well as the observed mass density profile of the OH GCS.
The best (in a least-squares sense) agreement with the observed 
profile is obtained when the density of the protogalactic 
cold gas is scaled downwards by a factor $2 \times 10^{-4}$.  $\chi ^2$ 
and Q values for the two bin sizes are provided in Table
\ref{tab:fit_rho_cold} (case a). 
Obviously, the shape of $\rho _{cold}(D)$ provides a good fit to
the observed distribution of the OH mass in our Galaxy.  The factor 
$2 \times 10^{-4}$ thus represents the formation efficiency of the halo 
bound stellar clusters which have managed to survive a Hubble time in 
the tidal fields of the Milky Way.  This factor of course represents 
a lower bound to the GC primordial formation efficiency.
As already highlighted by McLaughlin (1999), the next Section confirms
that this one is unlikely to have been exceedingly larger.  \\

Additionally, we note that the model, as described by equation 
(\ref{eq:rho_cold}), fits nicely the extent of the OH, as the cold 
gas density drops sharply at D $\simeq$ 40\,kpc.  The outer 
limit of the OH would thus arise from the depletion of cold gas at this
galactocentric distance.  However, we note that this agreement may be a 
mere coincidence only as it heavily depends on the choice of the 
parameters $\rho _{0hot}$, $\rho _{0DM}$ and $D_{DM}$ in equation 
(\ref{eq:rho_cold}).  Figure \ref{fig:rho_OH_modbis} illustrates how
variations in the adopted 
values of the parameters affect the shape of the cold gas profile.
The dashed-dotted and dotted curves represent equation (\ref{eq:rho_cold}) 
with the hot gas coefficient $\rho _{0hot}$ increased and reduced by 50 
per cent, respectively, while the dashed curve shows $\rho _{cold}(D)$ 
if the dark matter coefficients of Caldwell \& Ostriker (1981) are replaced 
by $\rho _{0DM}=30 \times 10^{6}\,{\rm M}_{\odot}.{\rm kpc}^{-3}$ and 
$D_{DM}=5$\,kpc (e.g., Baumgardt 1998).  In the last two cases, the gas 
density decreases smoothly with galactocentric distance, without
any sharp drop in the gas density.  In that case, the outer limit of 
the OH  might characterize a gas density threshold below which 
the formation of bound clusters is either inhibited or delayed.
In the first case, the gas density drops sharply at closer 
galactocentric distance (i.e., $D \simeq 25$\,kpc).
Nevertheless, we emphasize that all four curves fit well the 
observed OH profile in the range 1-20\,kpc, thus accounting for the vast 
majority of the OH GCs (less than ten per cent of the OH GCs are located
beyond $D \simeq 20$\,kpc).  This strengthens our 
hypothesis following which the mass density profile of the OH GCS
is tracing the cold baryonic material available to the
star formation process some 15\,Gyr ago.

\begin{figure}
\begin{center}
\epsfig{figure=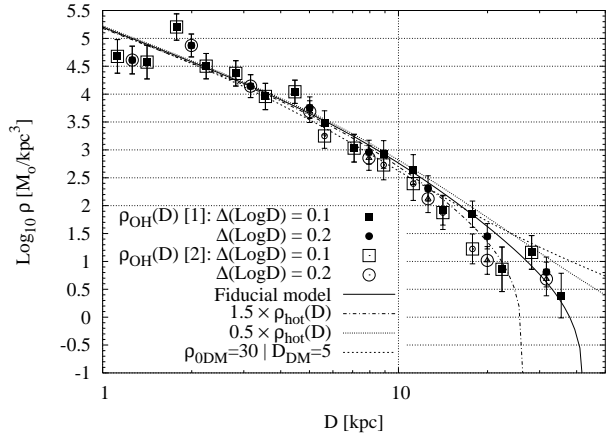, width=\linewidth}
\caption{Comparison between the same set of curves as in the previous figure
and the spatial distribution of the OH subsystem modified in the following 
way: exclusion from the initial 
sample of the GCs with extended structures (filled points) and
of the GCs with kinematics more typical of the YH (open points)} 
\label{fig:rho_OH_acc}
\end{center}
\end{figure}

An additional issue of potential concern is that a small, but
non negligeable, fraction of the OH may itself be an accreted component.
Several of the GCs in nearby dwarf galaxies have colour-magnitude diagram
characteristics that are indistinguishable from OH GCs (see, e.g., 
Grebel \& Gallagher 2004 and references therein).  Furthermore, 
Mackey \& Gilmore (2004) emphasize that, although the overwhelming
contribution of clusters from external galaxies would be to the Galactic YH
subsystem, there may also be a non-zero contribution to the OH subsystem,
which they estimate of order 15 per cent or, equivalently, a dozen of GCs.
Such accreted OH cluster candidates may be identified on the basis of 
either an extended core radius or spatial motions more typical of YH objects, 
although these alone are not definite indicators of an extra-Galactic origin.
Mackey \& Gilmore (2004) find five OH GCs in the first category 
(NGC~6809, 6101, 7492, 5897 and Pal~15) and 6 GCs in the second
(NGC~1904, 2298,5024,5904,6205,7089).  We have thus refitted equation   
(\ref{eq:rho_cold}) to the observed OH distribution, after rejecting
(1) the five GCs with extended structure,
(2) all eleven GCs quoted above plus Pal~2 (see below).  
These two observed spatial distributions are illustrated in
Fig.~\ref{fig:rho_OH_acc}.   
In the first case (filled symbols), one hardly detects a difference with
respect to the distribution of the whole OH GCS (as in
Figs.~\ref{fig:rho_OH_fits}-\ref{fig:rho_OH_modbis}).   
Accordingly, the parameters of the fit are much similar to previously
(compare case (a) and case (b) in Table \ref{tab:fit_rho_cold}).  
In a second step (open symbols in Fig.~\ref{fig:rho_OH_acc} and case
(c) in Table \ref{tab:fit_rho_cold}), 
we have removed the complete sample of 11 OH GCs suspected of having 
been accreted as well as Pal~2.  In the [Fe/H] vs HBR diagram, this one
is located at the frontier between the OH and the YH groups and
its nature is thus ill-determined. While we have previously considered 
it as an OH GC, we now sort it in the YH group. 
Removing the so-defined sample of 12 GCs, the changes 
are more significant but still, the modified observed distribution
is satisfactorily fitted by equation (8), the main point being that the model 
distribution overestimates the mass density of the OH subsystem
in the region $D \simeq 20$\,kpc.  We thus conclude that our fit is robust,
even though the actual radial distribution of the OH mass remains
slightly uncertain as a few old clusters may be accreted objects and, 
thus, interlopers with respect to the genuine initial Galactic GCS. 
 
In what follows, we consider the OH mass distribution displayed in 
Figs.~\ref{fig:rho_OH_fits} to \ref{fig:rho_OH_modbis} and equation
(8) as our fiducial observed and theoretical  
radial mass density profiles, respectively.

\begin{table}
\begin{center}
\caption[]{Results of fitting the spatial distribution of the baryonic
cold matter (equation 8) on the observed mass density profile of the OH 
subsystem: (a) all the GCs defined as OH objects on the basis of their
location in a [Fe/H] vs HBR diagram (Table \ref{tab:OH} and
Figs.~\ref{fig:rho_OH_fits}, 
\ref{fig:rho_OH_mod} and \ref{fig:rho_OH_modbis}); 
(b) same as (a) but excluding five GCs with extended structures
(filled symbols in Fig.~\ref{fig:rho_OH_acc}); (c) same as (a) but
excluding objects with  extended structure or extreme kinematics 
(open symbols in Fig.~\ref{fig:rho_OH_acc}) (see text for details)}
\label{tab:fit_rho_cold}
\begin{tabular}{ c c c r c } \hline 
     & $\Delta {\rm Log_{10}} D$ & $\Delta {\rm Log}_{10} \rho$ & $\chi ^2$ &  Q   \\ \hline
(a)  & 0.1                  &   -3.69 $\pm$ 0.06 & 15.0      & 0.45 \\ 
     & 0.2                  &   -3.70 $\pm$ 0.08 &  6.3      & 0.51 \\ \hline 
(b)  & 0.1                  &   -3.70 $\pm$ 0.06 & 14.5      & 0.49 \\ 
     & 0.2                  &   -3.71 $\pm$ 0.08 &  6.3      & 0.50 \\ \hline
(c)  & 0.1                  &   -3.78 $\pm$ 0.07 & 20.6      & 0.11 \\ 
     & 0.2                  &   -3.81 $\pm$ 0.08 & 11.1      & 0.14 \\ \hline
\end{tabular}
\end{center}
\end{table}   

\section{Evolution of the spatial distribution of the OH mass}
\label{sec:evol_rho_D}

As announced in Section 2, we now put on a firmer foot the 
hypothesis following which the initial GCS mass density 
profile is reasonably approximated by what it presently is. \\ 

Evolutionary processes act more efficiently upon low-mass GCs as well as 
on GCs located in the inner Galactic regions.  In other words, 
the evolution with time of the GCS has been mostly 
determined by the initial spatial distribution of the clusters in the 
Galactic halo as well as by their initial mass spectrum.  
In this respect, it is interesting to note that the sharp contrast between 
the fraction of surviving clusters and the ratio of the final to the 
initial total mass in clusters as obtained by McLaughlin 
(1999) (see Section 2) arises from the choice of a power-law mass
spectrum with a steep (i.e., --2) slope and probing down to very low-mass 
(i.e., of order $10^3\,{\rm M}_{\odot}$).  Other mass 
spectra could thus lead to different results. 

\begin{table*}
\begin{center}
\caption[]{Results of our simulations for different GC initial mass 
functions (IMF, see text for details). F$_N$ and F$_M$ are the 
fraction of surviving GCs and the ratio of the final to the initial 
mass in GCs, respectively.  Also given are the results of fitting the 
radial distributions evolved during 15\,Gyr to the data for 
the two different binnings considered, that is, the vertical logarithmic
shift $\Delta {\rm Log}_{10} \rho$ required to obtain the best match 
between the model and the data and the corresponding $\chi ^2$ and 
Q($\nu /2$, $\chi ^2/2$) values}
\label{tab:fit_evol_GCS}
\begin{tabular}{ c l c c c c c c c c c c c c } \hline 
Run & GC IMF & ~~ & F$_N$ & F$_M$ & ~~ &    & $\Delta {\rm Log}_{10} D$=0.1 &     &  &          & $\Delta {\rm Log}_{10} D$=0.2 &  \\ \hline
 & &  & &  & & $\Delta {\rm Log}_{10} \rho$ & $\chi ^2$ & Q & & $\Delta {\rm Log}_{10} \rho$ & $\chi ^2$ & Q  \\ \hline

[1] & Gaussian (Mean=5.03; $\sigma$=0.66) & & 0.75 & 0.66 & & -2.11 & 20.7 & 0.15 & &-2.14 & \,\,7.1 & 0.40 \\

[2] & Power-law (m$_{low}$=10$^3$~M$_{\odot}$) & & 0.06 & 0.39 & & -0.54 & 27.3 & 0.03 & & -0.56 & 10.9 & 0.14 \\

[3] & Power-law (m$_{low}$=10$^4$~M$_{\odot}$ )  & & 0.46 & 0.51 & & -1.44 & 27.2 & 0.03 & & -1.46 & 10.9 & 0.14 \\

[4] & Power-law (m$_{low}$=10$^5$~M$_{\odot}$)&  & 0.96 & 0.68 & & -2.30 & 20.4 & 0.16 & & -2.31 & \,\,6.9 & 0.44 \\ \hline

\end{tabular}
\end{center}
\end{table*}   

How the initial mass spectrum of the Galactic halo GCs looked like
is a much debated issue.  As mentioned earlier, the observed luminosity 
spectrum ${\rm d} N/{\rm d} L$ of numerous cluster systems recently formed
in starburst and merging galaxies is a power-law.  However, whether
the luminosity spectrum ${\rm d} N/{\rm d} L$ constitutes a faithfull mirror 
of the underlying mass spectrum ${\rm d} N/{\rm d} m$  is questionable.  
Both shapes agree only if the variations of the mass-to-light ratio from 
cluster to cluster are limited (as is roughly the case for the 
Galactic halo GCS).
Such a requirement may not be true for ongoing or recent starbursts:
the formation duration of such systems may be a significant fraction 
of the system's median age and, thus, age spread effects among 
the young star cluster population may not be negligible. 
Being an age related quantity, the mass-to-light ratio
can no longer be considered as a constant and the shape of the luminosity 
spectrum may differ substantially from the shape of the mass spectrum.  
Raising this issue, Meurer et al.~(1995) and Fritze v. Alvensleben (1998, 1999)
showed that systems of young GCs could display a power-law luminosity 
spectrum while the underlying mass spectrum is a broken power-law 
or, equivalently, a gaussian when the binning is logarithmic 
(${\rm d} N/{\rm d log} m$).  
The reason for this is that, owing to the fading with time of the 
cluster luminosity, high mass
clusters can be observed over a wide range of ages, while low-mass ones 
are detectable at young ages only.  As a result, low-mass 
clusters are underrepresented in the observed mass spectrum.  Thus, this 
one can be described by a two-index power-law (or equivalently a 
bell-shaped mass function) as it does not raise in 
the low-mass regime as steeply as in the high-mass regime.

Vesperini (1998) has indeed shown that the halo GCS could have started 
with a gaussian initial mass function.  Building on N-body simulations 
performed by Vesperini \&
Heggie (1997), he demonstrated the existence of a  
quasi-equilibrium GC mass function, that is, a mass function whose 
initial gaussian shape and parameters (mean and standard deviation)
are preserved during the entire evolution through a subtle balance
between disruption of clusters and evolution of the masses of those which
survived, even though a significant
fraction of the GCs is destroyed.  Interestingly, the mean and the 
standard deviation of this gaussian equilibrium mass function 
(${\rm log} (m/M_{\odot}) = 5.03$ and $\sigma = 0.66$) are remarkably close 
to those of the mass function of the halo GCS.  Thus, the initial GC 
mass function may have been very similar to what it is today.  
Obviously, the study of the temporal evolution of the GC mass spectrum
is not enough to unveil its initial shape as both a power-law and a 
gaussian mass functions evolve into the current GC mass function after a 
Hubble time.

If we {\it assume} that the halo GC initial mass spectrum was similar 
to what is observed today in 
starbursts and mergers, to derive the latter, one should
estimate the individual ages for all star clusters.  This can be done 
by, for instance, comparing the observed broad-band photometry to the 
colours generated by spectral evolution synthesis models
(see, e.g., Parmentier, de Grijs \& Gilmore 2003).  These models
will then provide an estimate of the mass-to-light ratio for each cluster
as a function of age, eventually enabling one to derive the intrinsic 
mass spectrum underlying the observed luminosity spectrum. 
Unfortunately, even such studies lead to unconclusive results.  
Analysing the system of young GCs in the Antenna 
merger NGC~4038/39, Zhang \& Fall (2001) show that the very young clusters
(i.e., with age less than 150~Myr), those which have remained unaffected
by the various dynamical effects, are distributed in mass according
to a pure power-law with slope $-$2.  On the other hand, 
de Grijs et al.~(2003) derive a gaussian mass function 
${\rm d} N/{\rm dlog} m$ (or equivalently a two-index power-law when 
binning the data in $m$) for an equally young system in 
the nearby starburst galaxy NGC~3310.

Therefore, at the present stage, neither the theory of the dynamical
evolution of a GCS, nor the observations of young GCs in starburst 
galaxies can help distinguishing between a power-law or a gaussian 
mass function for the GC initial population.
Considering the very unclear issue of this debate, we have 
computed the temporal evolution of the spatial distribution of the GCS mass 
around the Galactic centre for various mass functions. \\

\begin{figure}
\begin{minipage}[b]{0.995\linewidth}
\begin{center}
\epsfig{figure=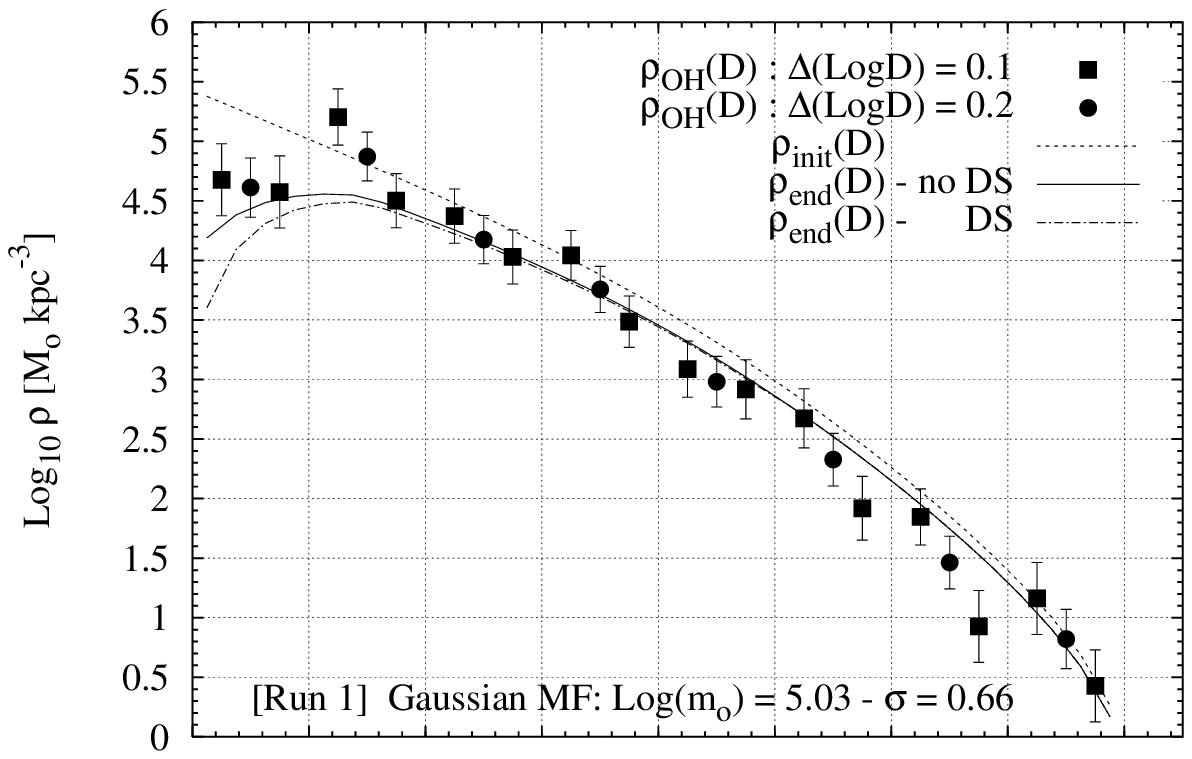, width=\linewidth}
\end{center}
\end{minipage}
\vfill
\vspace*{-8mm}
\begin{minipage}[b]{0.995\linewidth}
\begin{center}
\epsfig{figure=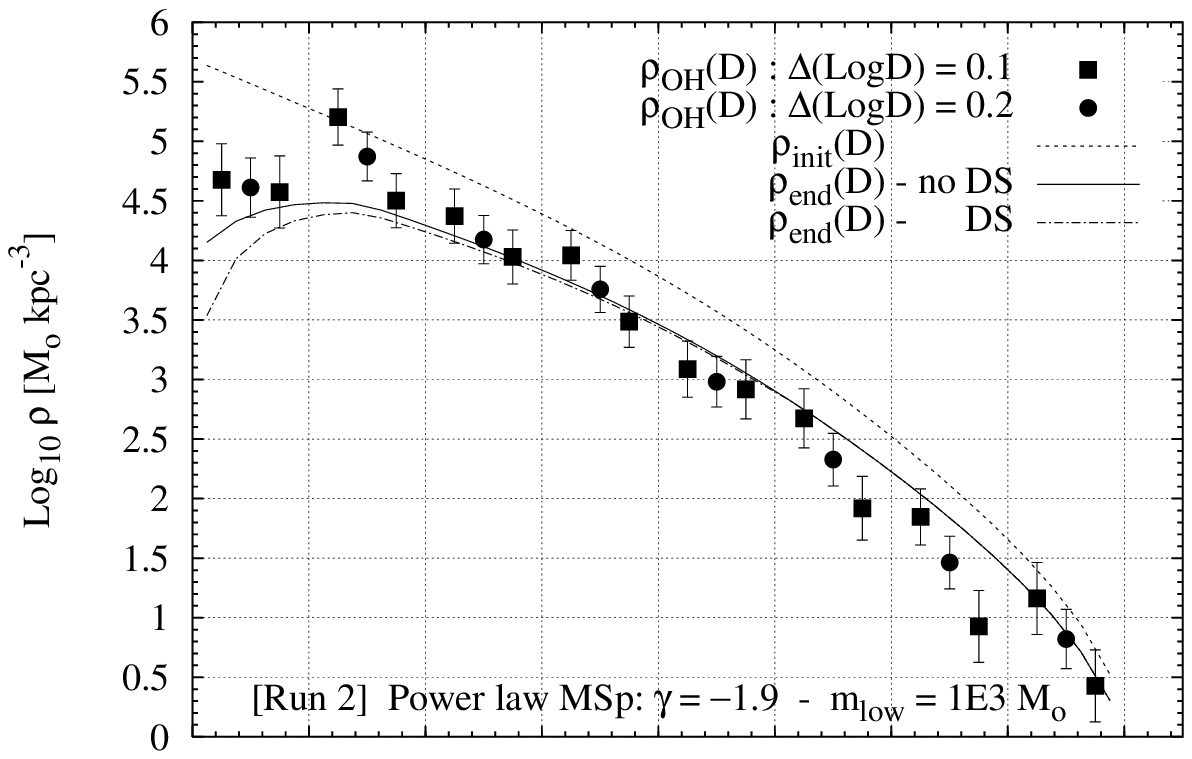, width=\linewidth}
\end{center}
\end{minipage}
\vfill
\vspace*{-8mm}
\begin{minipage}[b]{0.995\linewidth}
\begin{center}
\epsfig{figure=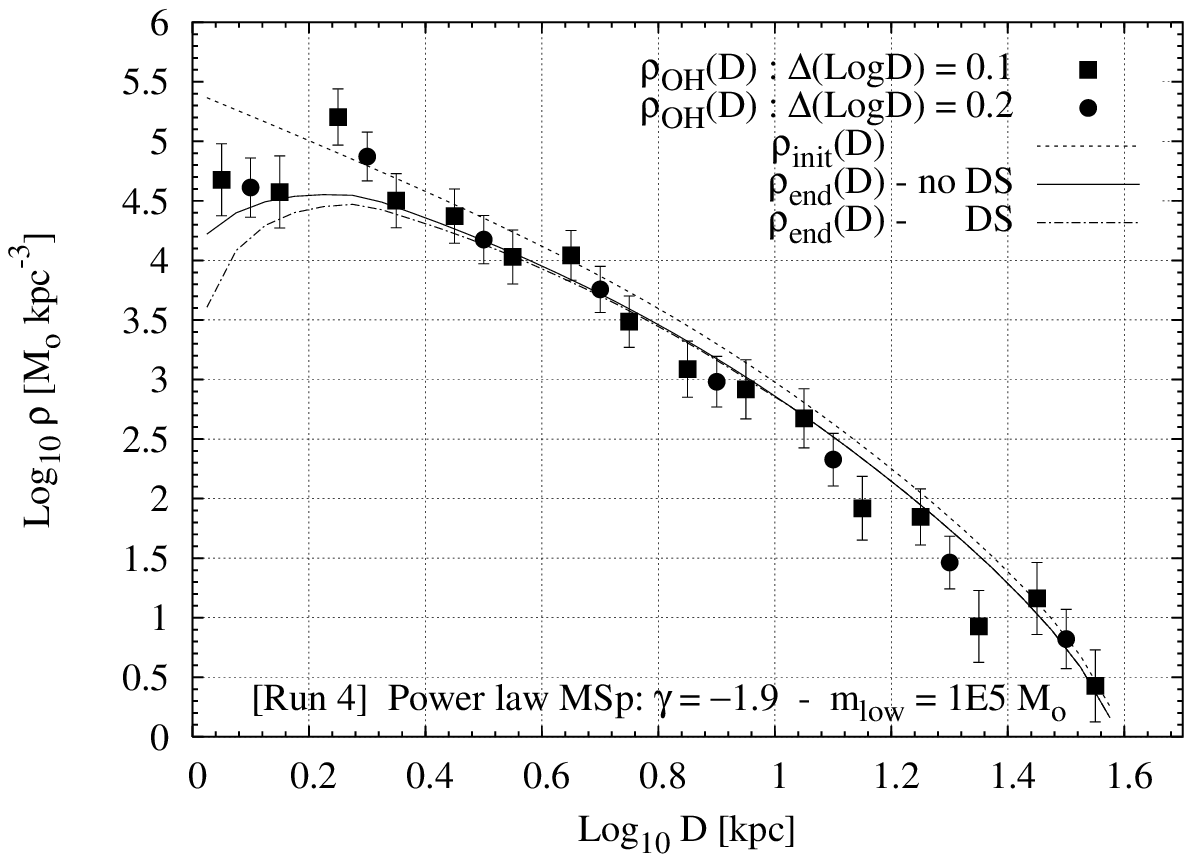, width=\linewidth}
\end{center}
\end{minipage}
\caption{Evolution with time of the radial mass density profile 
of a GCS whose mass is initially spatially distributed according to 
equation (\ref{eq:rho_cold}) (dashed curve).  Results with 
(dashed-dotted curve) 
and without (plain curve) disc-shocking (DS) are shown for three different 
initial GC mass spectra.  The plain curve is vertically 
shifted to match the observed data (see the quantity 
$\Delta {\rm Log}_{10} \rho$ in Table \ref{tab:fit_evol_GCS})}  
\label{fig:rho_evol}  
\end{figure}

To evolve the spatial distribution of the mass of a GCS from the time 
of its formation up to an age of 15\,Gyr, we adopt the analytic formulae 
of Vesperini \& Heggie (1997) which supply at any time $t$ the mass $m$ 
of a GC with initial mass $m_i$ and moving along a circular orbit at a 
galactocentric distance $D$.
These relations have been obtained by fitting 
the results of a large set of N-body simulations in which Vesperini \& 
Heggie (1997) take into account the effects of stellar evolution as well as
two-body relaxation, which leads to evaporation through the GC tidal 
boundary.  Disc shocking can also be included 
(see below).  In order to take into account dynamical friction,
GCs whose time-scale of orbital decay (see, e.g., Binney \& Tremaine 1987)
is smaller than $t$ are removed from the GCS at that time.  A summary of the
method is provided in Vesperini (1998, his Section 2). 

The temporal evolution of the mass of a GC, modelled as a multi-mass King 
model with an initial dimensionless concentration parameter $W_0 = 7$ and 
orbiting at constant galactocentric distance $D$, is assumed to follow:
\begin{equation}
\frac{m(t)}{m_i} = 1 - \frac{\Delta m_{st,ev}}{m_i}
- \frac{0.828}{F_{cw}} t\;.
\label{eq:mgc_t_noDS}
\end{equation} 

$\Delta m_{st,ev}/m_i$ is the fraction of cluster
mass lost due to stellar evolution (18 per cent in this particular model).
The time $t$ is expressed in units of 1\,Myr and $F_{cw}$, a quantity 
proportional to the initial relaxation time, is defined as:
\begin{equation}
F_{cw} = \frac{m_i \times D}{{\rm ln} N}\,,
\label{eq:Fcw}
\end{equation}

where $m_i$ and $D$ are in units of 1\,M$_{\odot}$ and 1\,kpc,
respectively, and $N$ is the initial number of stars in the GC.
To take into account disc shocking, the factor 0.828/$F_{cw}$ is merely 
replaced by $\lambda$ as defined by equation (3) of Vesperini (1998). \\  

We have distributed 20,000 GCs following a radial distribution obeying 
our equation (\ref{eq:rho_cold}) and various mass spectra:
\begin{itemize}
\item[a) ] a gaussian mass function ${\rm d}N/{\rm dlog}m$ with parameters 
equal to those of the equilibrium mass function of Vesperini (1998),
\item[b) ] three power-law mass spectra ${\rm d}N/{\rm d}m$ with a single 
slope of --1.9 and different lower mass limits, namely, 1E3, 1E4 
and 1E5\,M$_{\odot}$.
\end{itemize}
As for the last cut-off value, Fall \& Zhang (2001) indeed showed that the 
GC mass spectrum might have started with a truncation at mass of order 
1E5\,M$_{\odot}$, the low-mass tail of the currently observed GC mass 
distribution being formed as a result of the evaporation of the massive 
GCs located at short distance from the Galactic centre.  \\
 
Each panel of Fig.~\ref{fig:rho_evol} displays the results of one of 
these simulations (the case of a power-law truncated at 1E4\,M$_{\odot}$
is not represented as it is highly similar to the 1E3\,M$_{\odot}$ case).
The dashed, plain and dashed-dotted lines are, respectively, the initial
mass density profile as described by equation (\ref{eq:rho_cold}), the 
radial mass density profile after a Hubble time long evolution
without disc shocking (obtained through equation \ref{eq:mgc_t_noDS}) and 
with disc shocking.  The final profile without disc shocking (plain curve)
has been vertically shifted to provide the best agreement, in the 
least-squares sense, with the observed spatial distribution of the OH mass.
The amplitude $\Delta {\rm Log}_{10} \rho$ of the shift, the 
$\chi ^2$ and Q values of each fit, as well as the fraction of 
surviving clusters (F$_N$) and the ratio of the final to the initial total  
mass in GCs (F$_M$), are listed in Table \ref{tab:fit_evol_GCS}.  
F$_N$ exhibits an extremely large range of variation, with more than 90 
per cent of the clusters being disrupted if the initial mass spectrum
 is a single power-law going 
down to 1000 M$_{\odot}$, while a cut-off at 10$^5$ M$_{\odot}$ leads to 
the preservation of most of the GCs.  This large scatter in F$_N$ 
contrasts markedly with the limited range in F$_M$, as this one varies 
by less than a factor 2.  Therefore, considering our ignorance of the 
shape of the GC initial distribution in mass, mass-related quantities 
appear as indicators of the initial conditions more reliable than 
number-related quantities.  In the frame of the initial density profile
described by equation (8),
we also note that the initial mass of the OH GCS was at most 2.5 times
larger than what it currently is, thus suggesting that the bound
cluster formation efficiency in the Galactic halo was at most of order 
$2.5 \times (2 \times 10^{-4}$).  This result also agrees with previous 
studies (e.g., Baumgardt 1998, Vesperini 1998, McLaughlin 1999)
following which the disruption of GCs cannot be a major contributor 
to the build-up of the stellar halo.  
Indeed, assuming a mass-to-light ratio of 2.35 (see Section 2),
the mass of the OH GC subsystem is $\simeq 2 \times 10^7$M\,$_{\odot}$,
about two orders of magnitude less massive than the stellar halo 
($\simeq$ 10$^9$\,M$_{\odot}$, Freeman \& Bland-Hawthorn 2002). \\   

As already quoted, the shapes of the initial and final mass profiles 
are very similar, at least beyond a galactocentric distance of 
$\simeq 2$\,kpc.  It may be worth remembering that about one fifth of 
the cluster mass decrease is due to mass-loss associated with stellar 
evolution, which is independent 
of galactocentric distance.  As a result, the shape of the radial 
profile is left unaffected by this gaseous mass loss taking place during
the very early evolution of the GCs. 
The initial shape of the model profile being only moderatly affected by 
dynamical evolution, it is not surprising that the observed OH profile 
(indicated by the filled squares or circles in Fig.~\ref{fig:rho_evol}, 
depending on the size of the bins) and the outcome
of our simulations (plain curve in Fig.~\ref{fig:rho_evol}) are in
satisfactory agreement as 
indicated by the Q values listed in Table \ref{tab:fit_evol_GCS}.     
The agreement is weaker than in Table \ref{tab:fit_rho_cold} however,
especially for the power-laws with a truncation at 1E3\,M$_{\odot}$
and 1E4\,M$_{\odot}$,  
but this effect is mostly driven by one bin only, at log$D = 0.25$.  
Should it be ignored, the agreement between the model and the observations 
would jump up to $Q \simeq 0.3$ in the least favourable cases.
This kick in the observed density profile arises from the presence, 
at almost the same galactocentric distance, of two of the handful of very 
massive ($M_v \simeq -9.2$) GCs, i.e., NGC~6266 and NGC~6273 at 
D=1.7 and 1.6\,kpc, respectively.

In case of disc shocking (dashed-dotted curves in
Fig.~\ref{fig:rho_evol}), the ratios  
F$_N$ and F$_M$ are almost unaffected, being reduced by 
4 per cent at most with respect to the simulations without disc shocking
(see Table \ref{tab:fit_evol_GCS}).  The shape of the density profile
is also preserved except  
at galactocentric distances shorter than 2\,kpc where the predicted 
cluster density is significantly affected compared to the case without 
disc shocking.  In these regions, the disc shocking model underestimates 
the observed mass density.  This discrepancy may be 
solved if these very inner halo regions are contaminated by the metal-poor
tail of the bulge GCS.  Alternatively, it may be that the high-density
high-pressure environment of these very inner Galactic regions promoted
the formation of GCs with an efficiency locally higher than in the overall
halo.  In such case, the original GCS mass density profile around the 
Galactic centre may have been steeper than the density profile
$\rho _{cold}$. \\  

It is interesting to compare the values $F_N$ and $F_M$ listed in
Table \ref{tab:fit_evol_GCS} with those  
obtained in previous studies.  Considering the case of the equilibrium GC
mass function (run [1] in Table \ref{tab:fit_evol_GCS}), Vesperini
(1998) derived  
F$_N \simeq 0.55$ and F$_M \simeq 0.40$.  As for a power-law mass spectrum
with a slope --2 and a low-mass cut-off $\sim$1000\,M$_{\odot}$
(run [2] in Table \ref{tab:fit_rho_cold}), Baumgardt (1998) indicates
that the fraction  
of surviving clusters may be of order 1 per cent, while the final mass 
in GCs would represent 15 to 30 per cent of the initial mass content.  
These results, although in rough agreement with ours, are systematically 
lower.  This is not surprising however as these studies assume an initial
radial distribution scaling as D$^{-3.5}$ (even D$^{-4.5}$ for some of 
Baumgardt's runs), thus significantly steeper in the inner regions 
than our radial profile $\rho _{cold}$ (see Fig.~\ref{fig:rho_OH_mod}).  
Therefore, with respect to our 
simulations, a larger fraction of the initial population of GCs is 
put on orbits closer to the Galactocentric centre where they are more 
efficiently evaporated and eventually disrupted. \\

\section{Discussion and Summary}
We have proposed a model accounting for the spatial distribution
of the mass of the OH GCS.  We suggest that this GC subsystem
formed out of the cold baryonic gas present in the Galaxy some 15\,Gyr
ago, that is the gas leftover once the contribution of the dark matter is
removed, with a roughly constant formation efficiency throughout the halo. 
In order to test our hypothesis, we have built the presently 
observed radial distribution of the OH GCS, making use of the
latest classification of the halo GCs between the OH and YH 
subsystems (Harris 1996, updated 2003, Mackey \& Gilmore 2004).
We have found good agreement between our suggested model (equation 8)
and the observed OH mass distribution (see Fig.~\ref{fig:rho_OH_mod}
and Table \ref{tab:fit_rho_cold}).    
The fit is robust with respect to uncertainties in the observed 
distribution of the mass (i.e., depending on whether a few OH GCs may be 
accreted objects as the YH GCs) as well as with respect to the uncertainties 
in the parameters involved in equation (8). 

We have made one more step, demonstrating that the GCS mass density
profile, as described by equation (8), has not been severly affected 
by a Hubble time-long evolution in the tidal fields of the Milky Way, 
at least beyond galactocentric distances of order 2 to 3\,kpc.
This result is only weakly dependent on the shape of the initial GC mass 
spectrum, which is still very poorly known.  In fact, numerous studies 
found that both an initial gaussian mass function (i.e., a two-index 
power law mass spectrum) and an initial power-law mass spectrum will 
evolve into the presently observed gaussian mass function of GCs.  
Therefore, to recover the faint end of the original mass function 
of GCs on the basis of its temporal evolution only is not 
feasible.  We have thus considered various mass functions in order 
to demonstrate the robustness of the GCS mass density profile with
respect to the dynamical evolution.  

Our simulations make use of the analytic formulae obtained by 
Vesperini \& Heggie (1997), and the starting point consists in assuming
a given radial distribution and a given mass function for the GCs.  
Doing so, we neglect finer details such as the non-circularity of the GC 
orbits as well as the initial concentration of GCs.
As for the first point however, we recall that our main interest in this 
paper is the OH subsystem, which shows less extreme kinematics than the YH
GCs (see, e.g., Dinescu, Girard \& van Altena 1998).  
Thus, while the assumption of circular orbits for the GCs is clearly 
a simplifying one, this issue is likely to be less critical
than if we have dealt with the whole halo GCS or with the YH only.  
Regarding the second point, we note that the equation (9), which describes 
the temporal decrease of the cluster mass according to  
Vesperini \& Heggie (1997), implicitly assumes an initial concentration 
$W_0 = 7$ for all GCs.  The existence of GCs with lower initial 
concentration is thus clearly neglected.  Such low concentration 
GCs may have been initially present and destroyed 
relatively quickly by a combination of stellar winds and tidal limitation
with little or no help from two-body relaxation or gravitational shocks.   
In the inner Galactic regions (say within the Solar Circle), that is, where 
most of the GCS mass is located, even high-mass
GCs are unable to survive if their initial concentration is low
(Vesperini 1997).   By assuming that all GCs start with a concentration
$W_0 = 7$, we may thus overestimate the survival capacity of some 
high mass GCs and, thus, underestimate the decrease with time of
$\rho _{OH}(D)$.  However, it is worth pointing out that GCs
exhibit a correlation between their luminous mass estimate and their 
concentration in the sense that more massive GCs are more concentrated
(van den Bergh 1994).  This correlation is likely to be of 
primordial origin as it is stronger at larger galactocentric distances 
(Bellazini et al.~1996), that is, where evolutionary processes act on 
longer time-scales and, thus, where memory of the initial conditions 
is better preserved.  Moreover, simulations show that high-mass 
low-concentration GCs are destroyed in the inner halo but manage to 
survive at large galactocentric distances (i.e., $D \gtrsim 10$\,kpc, 
Vesperini 1997).  As they are not observed even beyond the Solar Circle,
high-mass low-concentration GCs were probably not part of 
the initial GCS and, therefore, to neglect them should not affect the results 
displayed in Fig.~\ref{fig:rho_evol} and Table \ref{tab:fit_evol_GCS}.   

Bearing these caveats in mind, we have shown that the shape of the evolved
radial mass density profile still nicely mirrors the initial one,
even though a significant fraction of the GCs has been destroyed
during a Hubble time long evolution.  Not surprisingly thus, the 
observed spatial distribution of the OH mass and the 
distributions arising out of our simulations are also found to be in good 
agreement (see Fig.~\ref{fig:rho_evol} and Table
\ref{tab:fit_evol_GCS}).  This strengthens 
our hypothesis following which the distribution of the gas GCs formed from
obeys equation (8) and, also, that the dark matter, whatever its nature, 
was not involved in the earliest stages of star formation.
 
This good agreement also suggests that the flattening of the observed
distribution of GCs around the Galactic centre may not result from 
tidal disruption processes only but may be, at least partly, of primordial
origin.  Indeed, in the region 1 to 3\,kpc, our suggested cold matter 
profile shows a flattening (slope $\simeq -2$) which contrasts with the 
overall shape, reasonably approximated by a slope of
$\simeq -3.5$ (see Fig.~\ref{fig:rho_OH_mod}).  The situation is still
unclear however as, at an age of 15\,Gyr, our mass distribution  
underestimates the observed mass density profile of the GCS for 
$D \lesssim 2$\,kpc, especially when disc shocking is included in the
simulations (see Fig.~\ref{fig:rho_evol}).  Possible solutions  
of this discrepancy include the contamination of the observed distribution 
by the metal-poor tail of the bulge/disc system and/or star formation 
efficiencies higher in the inner regions of the Galaxy owing to higher
pressures and densities.  As a result of this latter suggestion, the initial 
mass distribution of the GCs in the Galactic central regions may have been 
steeper than indicated by equation (8). 
Apart from these uncertainties, the satisfactory agreement between the model
and the observed radial density profile of the OH subsystem, beyond $D
\simeq$ 2 to 3\,kpc, suggests that the formation of GCs took place with a 
formation efficiency almost constant throughout the OH. 

\bsp


\label{lastpage}

\end{document}